\documentstyle[12pt,a41,epsfig,wrapfig]{article}

\newcommand{\lsim}{\raisebox{-0.07cm}{$\, \stackrel{<}{{\scriptstyle
\sim}}\, $}}

\def \bfk{\mbox{\boldmath $k_\perp$}}
\begin{document}

\thispagestyle{empty}

\vspace*{3cm}

\begin{center}
{\LARGE \bf A Possible Fixed Target Programme \\
\vspace*{+2mm}
      for the Polarized HERA Proton Ring\footnote{Contribution to the 
Proceedings of the 1997 Workshop on `Physics with Polarized Protons at HERA',\\
 DESY-Hamburg, DESY-Zeuthen, March-September 1997.} } 

\vspace{1cm}
                                         
{ V.A. Korotkov$^{a,b}$, W.-D. Nowak$^a$}\\

\vspace*{1.0cm}
                                         
{\it $^a$ DESY-IfH Zeuthen, D-15735 Zeuthen, Germany}\\

\vspace*{3mm}
{\it $^b$ IHEP, RU-142284 Protvino, Russia }\\

\vspace*{2.0cm}

\end{center}

\begin{abstract}
\noindent
The physics programme for a possible fixed target polarized nucleon-nucleon
collision experiment  aiming
at studying the nucleon spin structure at HERA is described.
The experiment named HERA-$\vec{N}$ could be realized using
an internal polarized gas target in the HERA polarized/unpolarized
proton beam. Single spin asymmetry measurements
could provide unique information on higher twist contributions.
Once the HERA proton beam is polarized measurements of the
polarized gluon  distribution using double 
longitudinal spin asymmetries in both photon and $J/\psi$ production
appear possible. A study of the Drell-Yan process for different
relative orientation of beam and target polarization 
can provide information on a variety of still not or badly
measured polarized structure functions.
The experiment would constitute a fixed target complement to the RHIC 
spin physics program with competitive statistical accuracy.

\end{abstract}

\vfill

\newpage

\setcounter{page}{1}

\begin{center}
{\LARGE \bf A Possible Fixed Target Programme \\
\vspace*{+2mm}
      for the Polarized HERA Proton Ring} 

\vspace{1cm}
                                         
{ V.A. Korotkov$^{a,b}$, W.-D. Nowak$^a$}\\

\vspace*{0.7cm}
                                         
{\it $^a$ DESY-IfH Zeuthen, D-15735 Zeuthen, Germany}\\

\vspace*{3mm}
{\it $^b$ IHEP, RU-142284 Protvino, Russia }\\

\vspace*{1.5cm}

\end{center}

\begin{abstract}
\noindent
The physics programme for a possible fixed target polarized nucleon-nucleon
collision experiment  aiming
at studying the nucleon spin structure at HERA is described.
The experiment named HERA-$\vec{N}$ could be realized using
an internal polarized gas target in the HERA polarized/unpolarized
proton beam. Single spin asymmetry measurements
could provide unique information on higher twist contributions.
Once the HERA proton beam is polarized measurements of the
polarized gluon  distribution using double 
longitudinal spin asymmetries in both photon and $J/\psi$ production
appear possible. A study of the Drell-Yan process for different
relative orientation of beam and target polarization 
can provide information on a variety of still not or badly
measured polarized structure functions.
The experiment would constitute a fixed target complement to the RHIC 
spin physics program with competitive statistical accuracy.

\end{abstract}
%

\section{Introduction}

\vspace{1mm}
\noindent
Past experiments studying collisions of polarized particles have reached quite
some progress, still the present picture of the nucleon spin structure
is essentially incomplete.
A number of new experiments at CERN, SLAC, DESY, BNL
was proposed to investigate in more detail polarized particle
interactions and to measure with good accuracy polarized
parton distributions. Up to now polarized deep inelastic
scattering experiments have measured only the longitudinal twist-2
spin structure function $g_1(x)$ and set an upper limit on the 
structure function $g_2(x)$ which contains a twist-3 contribution.
Polarized hadron-hadron experiments measured large single spin
asymmetries in inclusive production of different particles
($ \pi$, $\eta$, $K$, $p$, $\bar p$) and in elastic $pp$-scattering
which are all expected to be zero in perturbative QCD.
No measurements of the polarized gluon distribution, $\Delta G$,
and of chiral odd quark distributions have been accomplished up to now.

The experiment `HERA--$\vec{N}$' \cite{desy96-095}-\cite{97-004}
utilising an internal polarized nucleon target in the 820~GeV
HERA proton beam would constitute a natural extension of the studies
of the
nucleon spin structure in progress at DESY with the  HERMES
experiment \cite{her1}.
An internal polarized nucleon target offering unique features such as
polarization above 80\% and no or small dilution, can be safely
operated
in a proton ring at high densities up to $10^{14}$ atoms/cm$^2$.
The estimate of the integrated luminosity which could be accumulated in
 the experiment is based upon realistic figures. 
For the average beam and target polarisation
$P_B = 0.6$ and $P_T = 0.8$ are assumed, respectively. A combined trigger  and 
reconstruction efficiency of $C \simeq 50\%$ is anticipated.
Using $\bar{I}_B = 80 \; \mbox{mA} = 0.5 \cdot 10^{18} \; s^{-1}$
for the average HERA proton beam current and a rather conservative
polarized target density of $n_T = 3 \cdot 10^{13}$ atoms/cm$^2$ 
the projected integrated luminosity becomes
${\cal{L}} \cdot T = 240 \; pb^{-1}$
when for the total running time $T$ an equivalent 
of $T = 1.6 \cdot 10^7 \;s$ 
is assumed. This corresponds to about 3 real years under
present HERA conditions. 
One may argue, however, that at the time the experiment would run ( $>$2005 )
 even 500 $pb^{-1}$ {\it per year} might 
presumably become a realistic figure \cite{96-128} and the luminosity to be
accumulated over the lifetime of the experiment might be considerably higher.

As long as the polarized target would be used in conjunction with the unpolarized
HERA proton beam, the physics programme of HERA-$\vec N$ could be started and
focused to measurements of single spin asymmetries. 
Once having available a polarized proton beam at HERA,
all combinations of beam and target polarization
($LL,~TT,~LT$) could be possible and correspondingly double spin asymmetries
$A_{LL}$, $A_{TT}$ and $A_{LT}$ would be accessible at HERA-$\vec N$.

We note that single spin asymmetries might also be investigated in the
polarized HERA proton beam using an unpolarized gas target. This approach
would have a number of advantages, e.g.: i) unpolarized targets can deliver
higher densities (up to a limit given by the lifetime of the stored proton beam),
ii) the fragmentation region of the polarized nucleon lies at much smaller
laboratory angles which allows to use a forward oriented spectrometer.
Another possibility is to get the single spin asymmetries as a by-product of
the doubly polarized transverse-transverse collisions study whose importance will
be shown in section 3. Generally speaking, the physics prospects considered
in the section on the single spin asymmetries are rather independent on
whether beam or target is polarized.

The sensitivities shown in the rest of the paper are all calculated based
upon the above derived,
rather conservative estimate of 240 $pb^{-1}$ for the expected luminosity.

This paper intends to present a summary of the activities
undertaken so far to later propose a physics programme for an experiment
with an internal polarized gas target in the (polarized)
HERA proton ring. 
Any discussion of a possible lay-out of the experiment and its location
in the HERA ring is beyond the scope of this paper.

\vspace*{-3mm}
\section{Single Spin Asymmetries}

\noindent
A study of single spin asymmetries in inclusive particle
production is considered now as a way to investigate
higher twist effects: there might be twist-3 dynamical contributions
or hard scattering higher twists; there might also be intrinsic $\bfk$
effects, both in the quark fragmentation process and in the quark
distribution functions. The contributions of the different effects
are process dependent and therefore a comparative study of single spin
asymmetries in different processes might be a unique way of
understanding the origin and the importance of higher twist
contributions in large $p_T$ inclusive production.\\
There exists no consistent theoretical understanding yet of the role of higher 
twist contributions in single spin asymmetries. Existing phenomenological models
predict a size of these asymmetries ranging from a few to tens of percent.
Higher twist contributions should die out 
with increasing $p_T$ and the asymmetries should approach zero as $A_N \sim m/p_T$.
There is, however, another approach in which single spin asymmetries are
associated with the manifestation of non-perturbative dynamics and
the asymmetries would be large even at high $p_T$ \cite{troshin}. 
Among them, instantons are becoming increasingly interesting as possible sources
for single spin asymmetries \cite{kolya}. \\
In the following we discuss the capability of HERA-$\vec N$ to
investigate single spin asymmetries. 

\pagebreak

\begin{wrapfigure}{l}{6.0cm}
\vspace*{-4mm}
\centering
\epsfig{figure=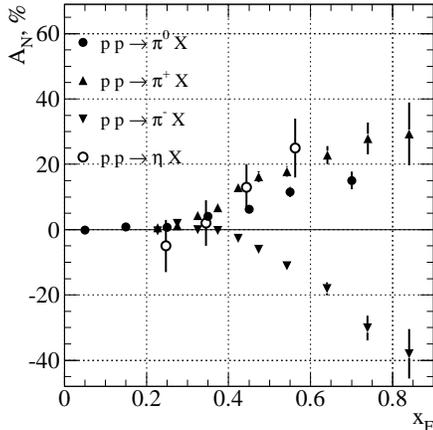,width=6.0cm}
\caption{\it  Single spin asymmetry in inclusive production of $\pi^{0\pm}$
         \protect \cite{704pi} and $\eta$ mesons \protect \cite{704eta}.
           }
  \label{fige704}
\end{wrapfigure}

{\bf Inclusive pion production} $p^{\uparrow} p \rightarrow
\pi^{0\pm}X$ exhibits surprisingly large single spin
asymmetries at large values of $x_F$, as it was measured a few years ago
by the E704 Collaboration using a transversely
polarized 200~GeV beam \cite{704pi}. For any kind of pions the asymmetry $A_N$
 (fig.~\ref{fige704}) shows 
a considerable rise above
$x_F > 0.3$, i.e. in the fragmentation region of
the polarized nucleon. It is positive for both
$\pi^+$ and $\pi^0$ mesons, while it has the opposite sign for $\pi^-$
mesons.
The charged pion data taken in the range $0.2 < p_T < 2$~GeV 
were split into two samples at $p_T$~=~0.7~ GeV/c; the observed rise
 is stronger for the high $p_T$ sample.\\
New results on the asymmetry in $\eta$ meson production were presented
recently \cite{704eta}. The asymmetry is positive and the behaviour is compatible
with the one observed in $\pi^0$ and $\pi^+$ production (cf. fig.\ref{fige704}). \\
There exist many results on asymmetry measurements in inclusive particle
production at smaller energies. Recently, a new experiment with a 40~GeV polarized
proton beam published data on the $p_T$ dependence in the range 
$0.7 \leq p_T \leq 3.4$ GeV/c, of the single spin asymmetry in
$\pi^{\pm}$, $K^{\pm}$, $p$ and $\bar p$ production in the central region
($0.02 \leq x_F \leq 0.10$) \cite{40gev}. 
The $p_T$ dependence measured for the $\pi^+$ asymmetry 
is compatible with older data obtained at beam energies 
of 13.3 and 18.5 GeV/c \cite{13gev} if plotted as a function of $x_T=2 p_T/\sqrt s$.
This appears to be in some contrast to the E704 data on the $\pi^0$ asymmetry 
in the central region which shows a result compatible with zero up to 
$p_T$ of about 4~GeV/c \cite{e704central}. 

\begin{figure}[hb]
\centering
\begin{minipage}[c]{5.5cm}
\centering
\epsfig{file=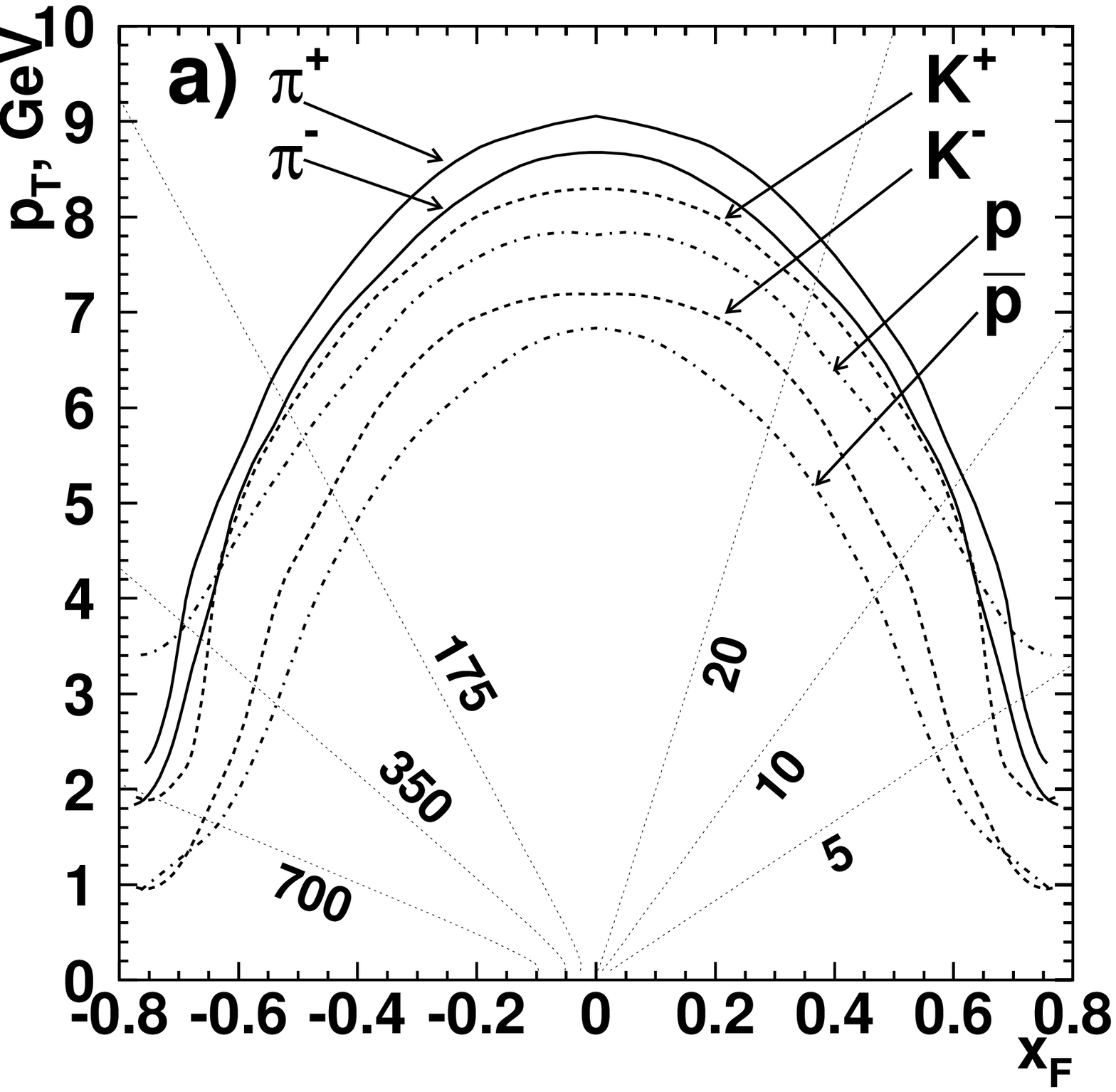, width=5.5cm}
\end{minipage}
\begin{minipage}[c]{5.5cm}
\centering
\epsfig{file=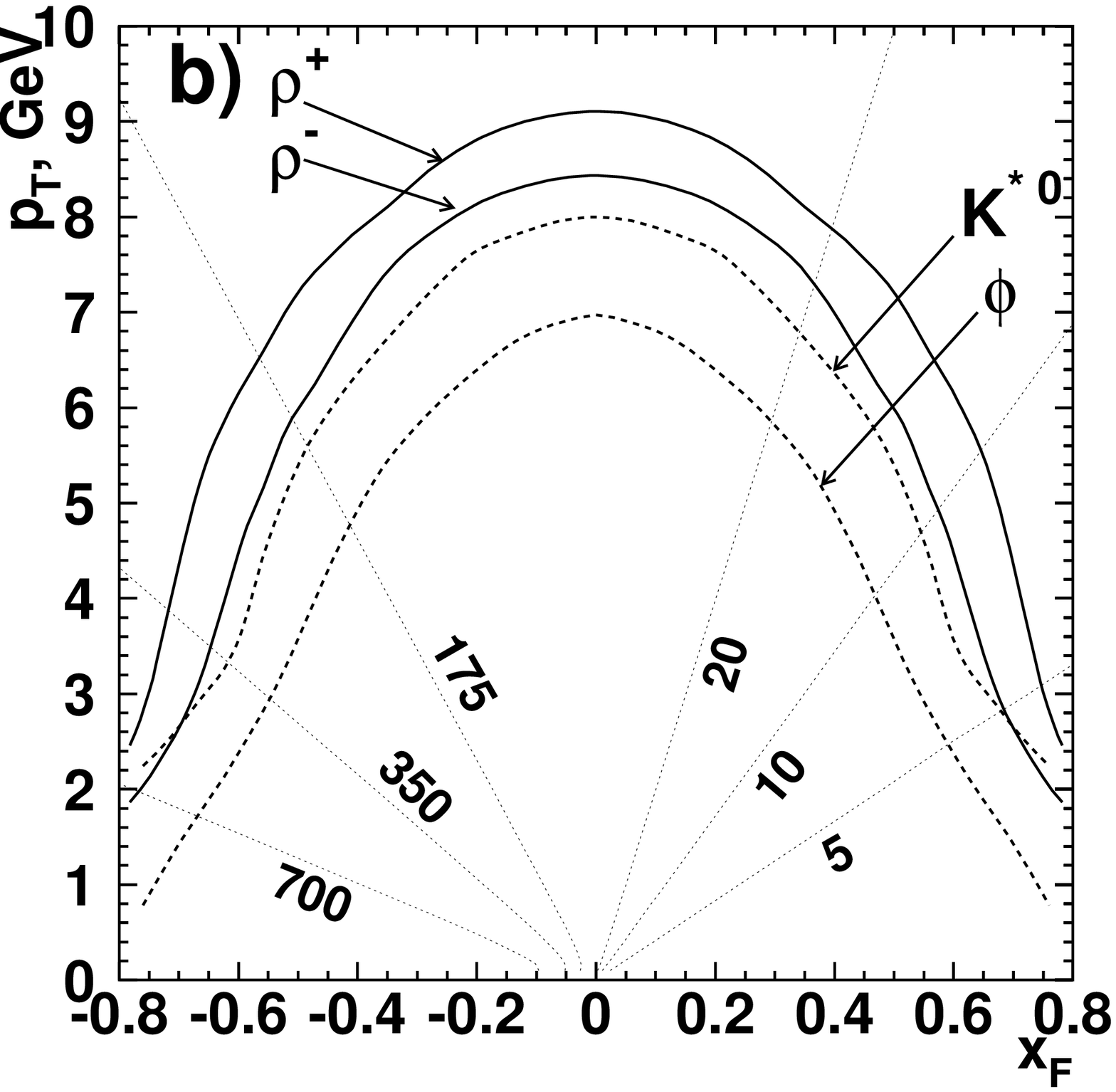, width=5.5cm}
\end{minipage}
\begin{minipage}[c]{5.5cm}
\centering
\epsfig{file=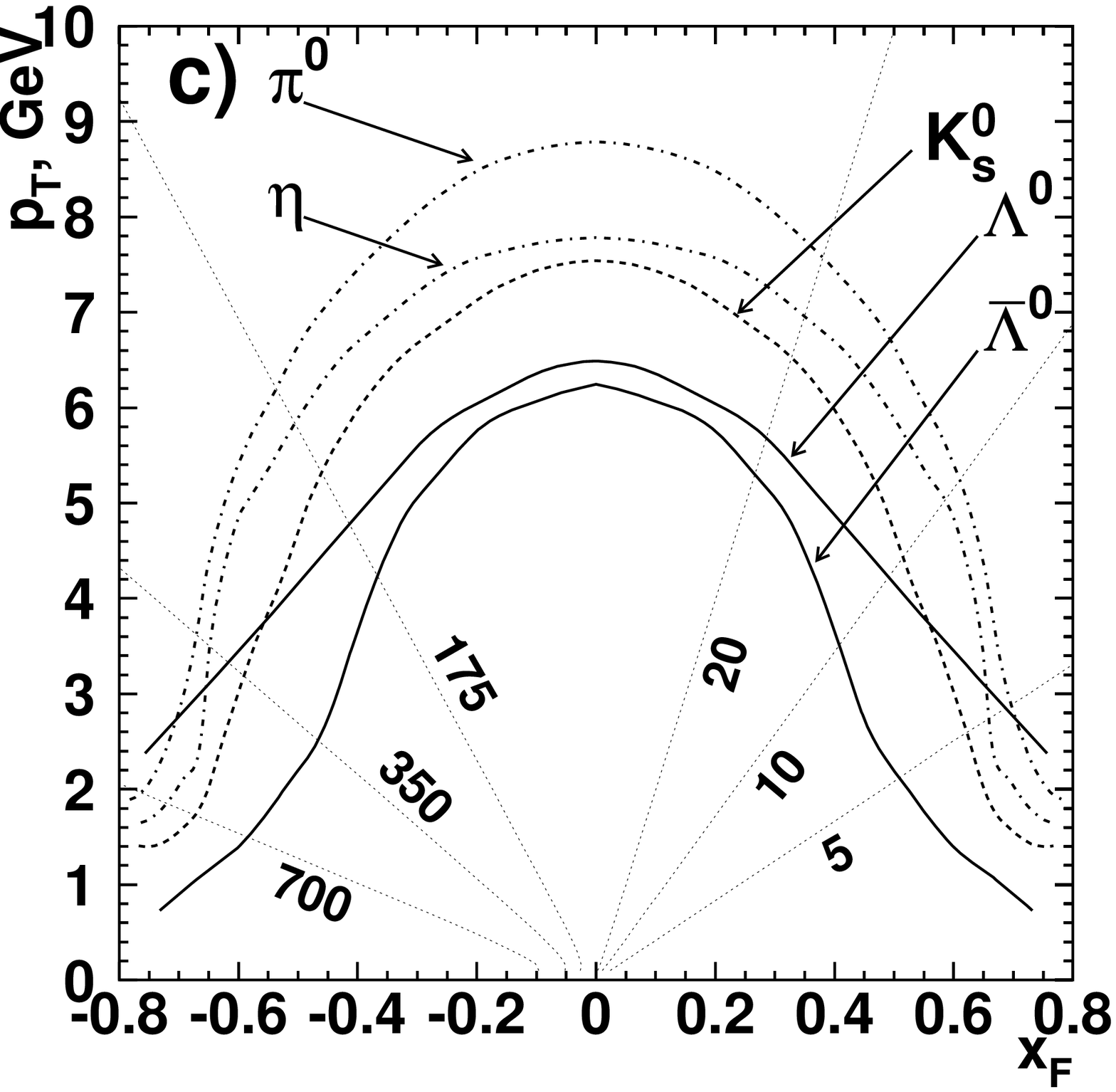, width=5.5cm}
\end{minipage}
\caption{\it Contours of the asymmetry sensitivity level $\delta A_N = 0.05$
             for inclusive production of different particles in the $(x_F, p_T)$
             plane. Only the following decay modes are taken into account:
             $K^{*0} \rightarrow K^+ \pi^-$, $\phi \rightarrow K^+ K^-$,
             $\eta \rightarrow \gamma \gamma$,  $K^{0}_s \rightarrow \pi^+ \pi^-$,
             $ \Lambda^0 \rightarrow p \pi^-$,
             $ \bar{\Lambda}^0 \rightarrow \bar p \pi^+$.
             Lines of constant laboratory angles of the particles are shown
             and marked with their values in units of mrad.
             }
\label{sensit}
\end{figure}

The $p_T$ values accessible with HERA-$\vec{N}$ would be significantly
larger than in all experiments performed up to now.
The sensitivity $\delta A_N$ of the asymmetry measurement in inclusive production
of different particles at HERA-$\vec N$ was calculated using the inclusive differential
cross-sections obtained with the Monte-Carlo program PYTHIA~5.6 \cite{pythia}.
The results are shown in fig.~\ref{sensit} in the ($x_F$, $p_T$) plane 
as contours characterizing the sensitivity level $\delta A_N = 0.05$
in a bin of $\Delta p_L \times \Delta p_T = 2 \times 2 $ (GeV/c)$^2$.
For produced particles lines of constant polar angle in the laboratory system
are shown; they are given for pions, but represent also a good approximation
for heavier particles. 

Experimentally, it is not a simple task to measure single spin asymmetries
 in the fragmentation
region of the polarized nucleon in a fixed target experiment at 820~GeV.
This region lies either at very 
large laboratory angles (a few tens of degrees) if a combination of polarized target
and unpolarized beam is used,
or it is at very small angles (a few mrad) 
for the other combination, unpolarized target and
 polarized beam (see fig.~\ref{sensit}).
The question how close to the HERA proton beam particles can be measured
deserves a special study. \\
As can be seen from fig.~\ref{sensit},
 the combined $p_T$ dependence of all involved higher-twist
 effects can be 
\begin{wrapfigure}{l}{7.0cm}
\vspace*{-8mm}
\centering
\epsfig{figure=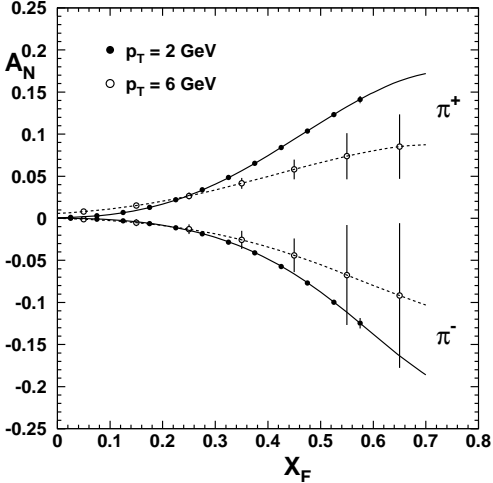,width=7.0cm}
\caption{\it Capability of HERA-$ \vec N $ to discriminate predictions
           \protect \cite{ans} for different $p_T$. }
             \label{asymurgia}
\vspace*{-5mm}
\end{wrapfigure}
measured with good accuracy ($\delta A_N \leq 0.05$)
up to transverse momenta of about 8$\div$10~GeV/c in the central region
$|x_F| < 0.2$ and up to 5$\div$6~GeV/c in the target fragmentation region.
Hence the $p_T$-range where higher twist effects are expected to be essential
 would be well covered.
The capability of HERA-$\vec{N}$ to really prove a $p_T$ dependence 
in the fragmentation region of the polarized nucleon
is shown in fig.~\ref{asymurgia}, where the curves were
obtained \cite{96-128}
assuming a non--zero quark distribution analysing power,
$N_q(\bfk)$, according to Ref. \cite{ans}. 
The curves and the projected statistical errors in fig.~\ref{asymurgia} 
are drawn for the 
combination of polarized proton beam and unpolarized target 
and the minimal pion detection angle was assumed to be 5~mrad. \\
A measurement of the asymmetry in the inclusive production of particles
with different quark contents  may allow to study a flavour dependence
of the higher twist contributions. In particular, it would be interesting to compare
the asymmetry for pions and kaons what appears possible at HERA-$\vec N$
(see fig.~\ref{sensit}a,c). \\

\vspace*{-0.6em}
A sizeable {\bf inclusive production of $\Lambda^0$ and $\bar{\Lambda}^0$ hyperons}
would allow to study the asymmetry in their production up to $p_T$ of about
 5$\div$6 GeV/c (fig.~\ref{sensit}c). The measurement of the final-state 
$\Lambda$ polarization via its decay would allow to study the polarization
spin transfer coefficient, $D_{NN}$. A recent study by E704 \cite{e704lambda}
at moderate values of $p_T$ (0.1$\div$1.5 GeV/c) showed a sizeable (up to 30\%)
spin transfer from the incident polarized proton to the outgoing $\Lambda^0$. \\

\vspace*{-0.6em}
The study of polarization asymmetries in {\bf inclusive 
vector meson production}
is especially attractive as these particles are produced 
`more directly' in comparison to pions which
are mainly decay products of heavier particles. 
Comparing asymmetries in vector and
pseudoscalar meson production can provide information on the 
magnitude of the asymmetry in quark scattering \cite{czyz}. 
If the asymmetry is generated only
during the fragmentation of polarized quarks,
the asymmetry of 
$\rho$ mesons is expected to be opposite 
in sign to that of pions,
$R_{\rho / \pi} = A_N^\rho / A_N^\pi \simeq -{1 \over 3}$.
On the contrary, if the quark scattering asymmetry  were 
the dominating one, the asymmetries of pseudoscalar and vector
mesons would not differ substantially. \\
The statistical sensitivity of HERA-$\vec N$ for measuring single spin asymmetries
in inclusive production of $\rho$, $K^{*0}$, and $\phi$
vector mesons are presented in fig.~\ref{sensit}b. The sensitivity
for $\rho$ production is at a level comparable to that for pions (fig.~\ref{sensit}a),
while for $K^{*0}$ and $\phi$ mesons the reachable $p_T$ values are lower.
On the other hand, a study of the asymmetry in $K^{*0}$ and $\phi$ production using
the decay channels $K^{*0} \rightarrow K^{\pm} \pi^{\mp}$ and $\phi \rightarrow
K^+ K^-$ could be easier since the level of the expected combinatorial 
background is smaller.
Also, the asymmetry in $\phi$ meson production could be useful 
for a study of the strange quark polarization in a nucleon \cite{troshin}. \\

{\bf Inclusive direct photon production}, $p p^{\uparrow} \to \gamma X$,
proceeds without fragmentation, i.e. the photon carries directly the
information from the hard scattering process. Hence this process
measures
a combination of initial $\bfk$ effects and hard scattering twist--3
processes. The first and only results up to now were obtained by the 
E704 Collaboration \cite{Phot704}
showing an asymmetry compatible with zero within
large errors for $2.5 < p_T <3.1$~GeV/c in the central region
$ | x_F | \lsim 0.15$.   \\
The experimental sensitivity of HERA-$\vec N$
was determined using cross-section calculations
 for the two dominant hard subprocesses, i.e. gluon--Compton scattering
($qg \rightarrow \gamma q$) and quark--antiquark annihilation
($q \bar q \rightarrow \gamma g$),
and of background photons that originate mainly from $\pi^0$ and
$\eta$ decays.
It turns out that a good sensitivity (about 0.05)
can be maintained up to $p_T \leq$ 8 GeV/c.
For increasing transverse momentum the annihilation subprocess and the
background photons are becoming less essential;
we expect to be able to detect a clear dependence on $p_T$,
of the direct photon single spin asymmetry. \\

There is an interesting possibility \cite{coll94,jaf97} to extract
the third twist-2 quark distribution function (quark
transversity distribution, $\delta q(x)$ or $h_1(x)$) 
using {\bf inclusive production of two pions} on the transversely polarized
nucleon, $p~+~p^{\uparrow}\rightarrow~ \pi^+~+\pi^-~+~X$.
This structure function describing basically the fraction of
transverse polarization of the proton carried by its quarks is
totally unknown at present.
In inclusive lepton DIS its contribution is suppressed 
by a quark mass whereas it is in principle accessible in 
semi-inclusive DIS \cite{her1,jaf97,JaffeJi1}. 
The asymmetry in inclusive two-pion production would be studied as a function
of the angle of the normal of the two-pion plane, $\vec{k}_+ \times
\vec{k}_-$, with respect to the polarization vector, $\vec{S}_\perp$,
of the nucleon. The statistical sensitivity of HERA-$\vec{N}$ remains to
be calculated. \\

The single spin asymmetries in {\bf inclusive J$/\psi$ production} 
\cite{desy96-04} and in {\bf Drell-Yan production},
$p~+~p^{\uparrow}\rightarrow~ l \bar l~+~X$, at small
transverse momenta \cite{hammon}, were 
estimated at HERA-$\vec N$ energy to be of the order of $0.01 \div 0.02$.
Nevertheless,  one may expect larger asymmetries as the calculations
might still not be complete. The projected level of sensitivity
can be taken from the section on double spin asymmetries as it is the same 
for both cases if the beam polarization is accounted for in case of $A_{LL}$. \\

Large spin effects in {\bf proton-proton elastic scattering},
$p~+~p^{\uparrow}\rightarrow~p~+~p$, have been disco\-vered many years ago.
The single spin asymmetry
$A_N$ was found significantly different from zero. 
At HERA-$\vec N$ energy one can measure the asymmetry in the range of
$p^2_T = 5 \div 12$ (GeV/c)$^2$ (see ref.\cite{desy96-04,96-128}). 


\section{Double Spin Asymmetries}

\vspace{1mm}
\noindent
The measurement of double spin asymmetries
in certain final states seems to be the most valuable tool to measure
polarized gluon and quark distribution
functions in the nucleon. The most accurate way to do so
is the study of those processes which can be calculated
in the framework of perturbative QCD.
Production of {\it direct photon (plus jet)},
{\it J/$\psi$ (plus jet)}, and {\it Drell-Yan pair} final states
are most suited because there are only small uncertainties due to
fragmentation. \\
In the following we discuss the capabilities of
HERA-$\vec{N} $, operated in doubly polarized mode,
to perform such measurements.

\subsection{${{\Delta G} \over {G}}(x)$ Measurement}

Prospects for the polarized gluon distribution measurements with
HERA-$\vec N$ are presented in other contributions
\cite{dGHERAN,jpsi-ntt} to these proceedings. Here only a brief summary is
given. \\

{\bf Direct photon production} in $pp$~interactions is dominated 
by the quark-gluon Compton subprocess, $q(x_1) + g(x_2) \rightarrow
\gamma + q$ and the asymmetry 
$A_{LL}$
 is directly sensitive to the polarized gluon distribution. 
Indeed, the NLO calculation \cite{gor1} of prospects for the
 {\bf inclusive photon production} study
with HERA-$\vec{N}$ showed a sufficient
statistical accuracy of HERA-$\vec{N}$
to discriminate between different polarized gluon distribution
functions. \\

The production of $c \bar c$ quarkonium
states, in particular {\bf inclusive $J/\psi$ production},
 is a similarly clean tool to
measure the polarized gluon distribution.
For the production of quarkonia with $p_T$ above 1.5~GeV 
the $2 \rightarrow 2$
subprocess $g(x_1) + g(x_2) \rightarrow (c \bar c) + g$ provides the main
contribution.
Because of the relatively large quark mass the $c\bar c$
production cross section and the expected asymmetry are supposed to be
calculable perturbatively.
The expected asymmetry is proportional to $(\Delta G(x) / G(x) )^2$ 
and the projected statistical accuracy of HERA-$\vec N$ allows
for a very good discrimination
between different parametrizations of $\Delta G(x)$ \cite{96-128}.

\begin{wrapfigure}{r}{7.0cm}
\vspace*{-7mm}
\centering
\epsfig{file=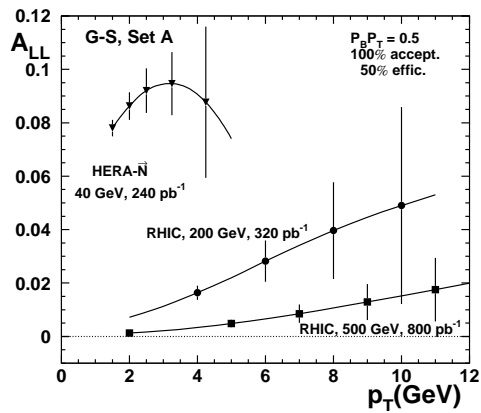, width=7.0cm}
\caption{\it The asymmetry and projected
statistical errors for $J/\psi$ production
at HERA-$\vec N$ and two different energies at RHIC as a function of $p_T$. }
\label{jpsirhic}
\end{wrapfigure}

The expected double spin
asymmetry for $J/\psi$   production at RHIC energies is much smaller.
In fig.~\ref{jpsirhic} predictions at HERA-$\vec N$ and 
two different RHIC energies are shown with projected statistical errors.
In the statistically accessible $p_T$ interval the asymmetry
ranges between 0.08 and 0.10 at HERA-$\vec N$, but only
between 0.01 and 0.03 at RHIC energies.
It is likely
that the fixed target experiment at HERA might accomplish a more
significant measurement of the charmonium production asymmetry. \\

The inclusive {\bf two-jet production} involves several hard subprocesses
($gg$, $gq$, $qq$ scattering) with gluons.
The extraction of the two-jet production cross-section at the given HERA-$\vec N$
fixed target kinematics is problematic, as was described in some more detail 
in ref.~\cite{desy96-04}. As a possible way out one may presumably
replace the two jets by two
correlated high $p_T$ hadrons opposite in azimuth.  \\
The asymmetry in {\bf open charm production} could possibly be
measured using as a tag a high $p_T$ single muon or electron-muon
pairs from charm decays. 
This option, as well as two-jet production, needs further study. \\

\begin{wrapfigure}{r}{9.0cm}
\vspace*{-7mm}
\centering
\epsfig{file=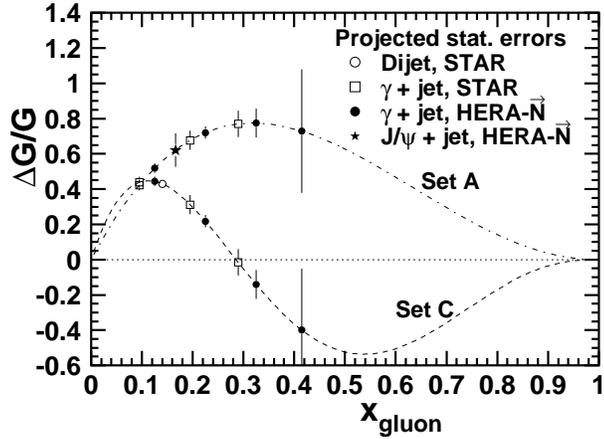,width=9.0cm}
\caption{\it Typical predictions for the polarized gluon
       distribution (LO calculations
       from Ref. \protect \cite{GehrStir}) confronted
       to the projected statistical errors expected for
       HERA-$\vec{N}$ and RHIC experiments.  }
      \label{gamjet}
\end{wrapfigure}

The complete kinematics of the underlying hard
2$\rightarrow$2 subprocess can be reconstructed if the away-side jet
in the production of photon or $J/\psi$ is measured, as well. In this
case the asymmetry $A_{LL}$ can be directly related to the polarized
gluon distribution \cite{desy96-04}.
Using this approach {\bf photon plus jet} production was discussed in
Ref. \cite{96-128} as a tool to directly measure $\Delta$G/G.
In fig.~\ref{gamjet} the projected statistical sensitivity of 
HERA-$\vec{N}$ for the
$\Delta G(x)/G(x)$ measurement,
on the present level of understanding, is shown vs. $x_{gluon}$ in
conjunction
with predicted errors for  STAR running
at RHIC at 200 GeV c.m. energy \cite{yok1}.
The errors demonstrate clearly that in the region
$0.1 \leq x_{g} \leq 0.4$ a significant result from {\it photon plus
jet production} can be expected from HERA-$\vec N$
 with an accuracy being about competitive
to that predicted for RHIC. \\

In {\bf $J/\psi$ plus jet} production the quark-gluon subprocess 
contributes only about 10\% to the asymmetry compared to the 
gluon-gluon fusion subprocess.
The prospect of a $\Delta G(x)/G(x)$ measurement at HERA-$\vec N$
is shown as an additional entry in fig.~\ref{gamjet}.
Although the $x_{gluon}$ interval ($0.1 \div 0.4$) explored by both
HERA-$\vec N$ and RHIC is quite comparable, the different 
transverse momentum ranges accessed (2...8 GeV at HERA-$\vec N$;
10...40 GeV at RHIC) make both measurements indeed complementary;
${\Delta G} \over G$ would be studied by HERA-$\vec N$ in the pQCD 
onset region whereas the RHIC experiments will explore 
${\Delta G} \over G$ in the deep perturbative region. \\
%


\subsection{Study of the Doubly Polarized Drell-Yan Process}

\vspace*{+0.5ex}
The production of {\bf Drell-Yan pairs} in polarized nucleon-nucleon 
collisions can provide information on a variety of polarized structure
functions in dependence on the relative orientation of the beam and
target polarization directions. \\

\vspace*{+0.5ex}
The {\bf longitudinal} double spin asymmetry
turns out to be well suited to extract the
polarized light sea-quark distribution
\begin{equation}
\label{DYLL}
A^{DY}_{LL} = - {{\sum_i e^2_i [\Delta q_i(x_1) \Delta \bar q_i(x_2) + (1
    \leftrightarrow 2)]} \over 
{\sum_i e^2_i [q_i(x_1) \bar q_i(x_2) + (1
  \leftrightarrow 2)]} } .
\end{equation}
\begin{figure}[htb]
\centering
\begin{minipage}[c]{8.0cm}
\centering
\epsfig{file=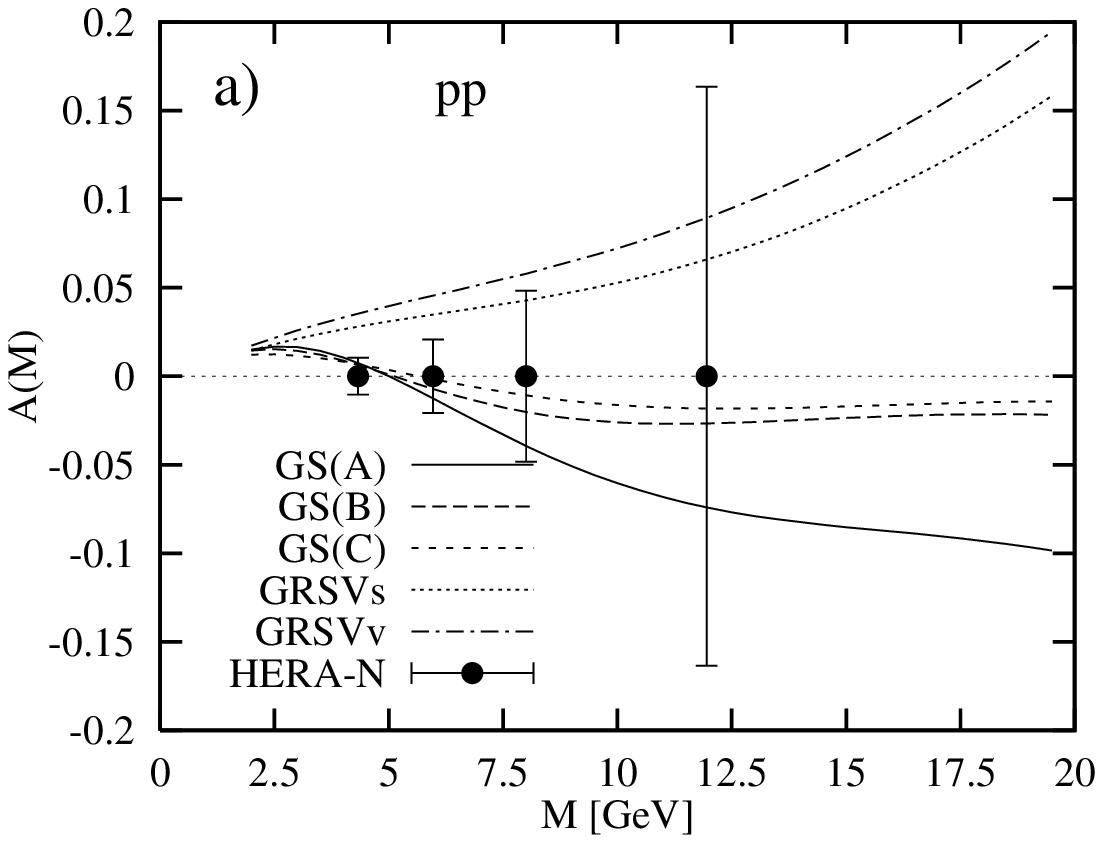, width=8.0cm}
\epsfig{file=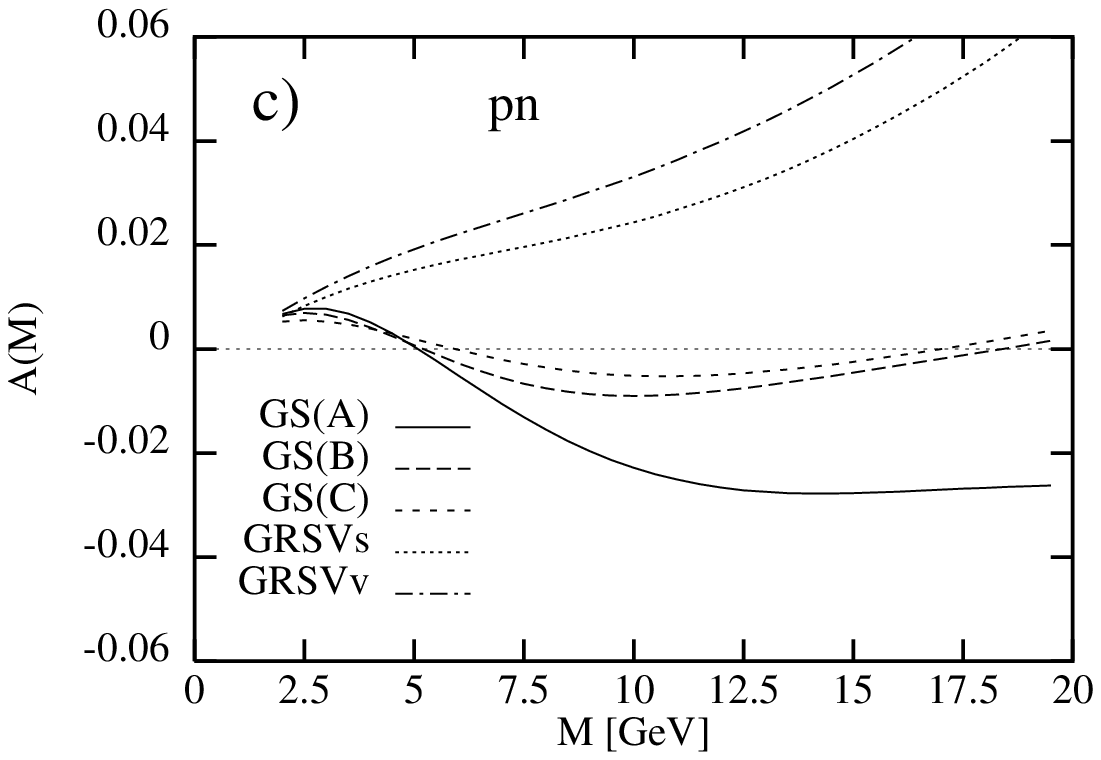, width=8.0cm}
\end{minipage}
\hspace*{0.5cm}
\begin{minipage}[c]{8.0cm}
\centering
\vspace*{1mm}
\hspace*{0.5cm} \epsfig{file=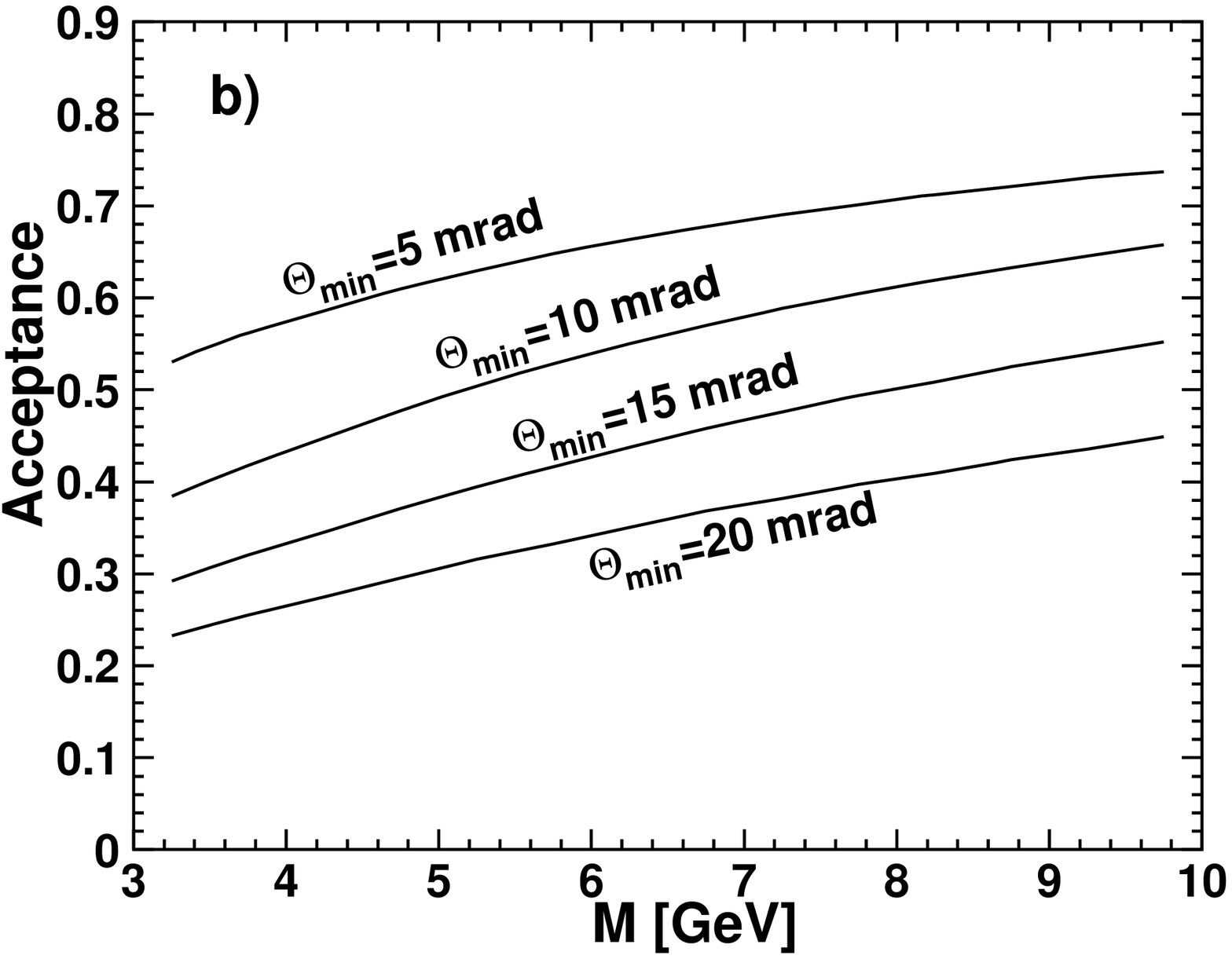, width=8.0cm}
\epsfig{file=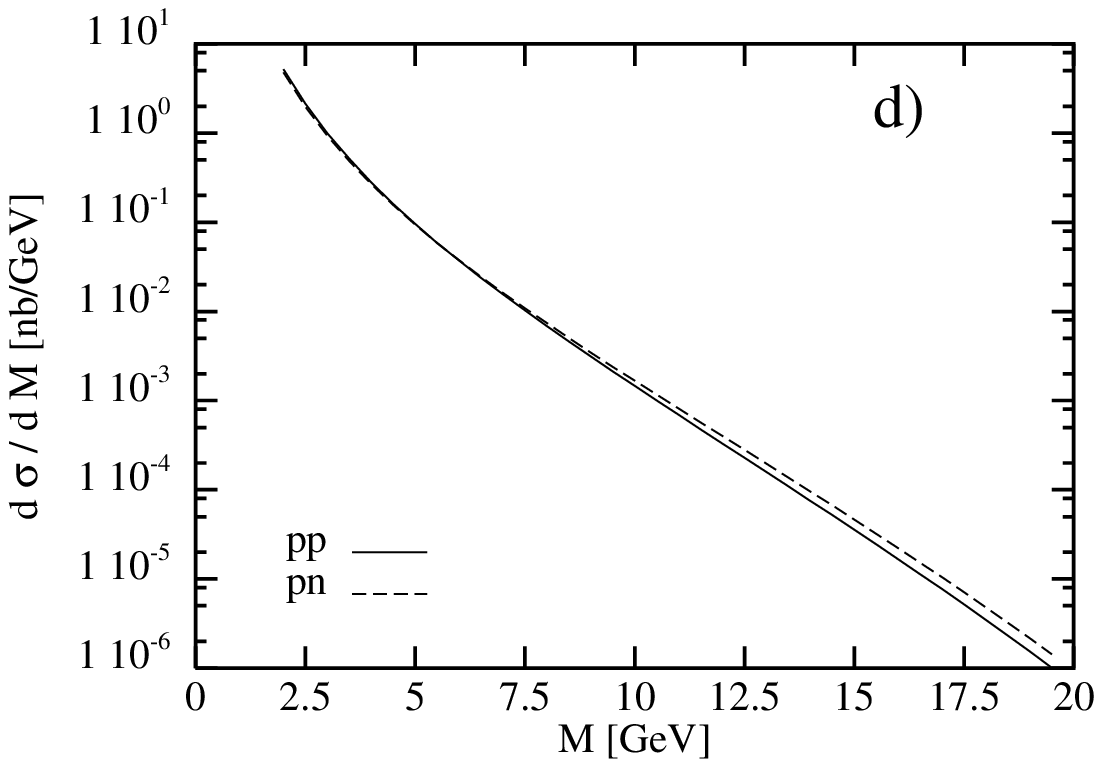, width=8.0cm}
\end{minipage}
\caption{\it  Expected longitudinal double spin asymmetries in the polarized
       Drell-Yan process for a) pp and c) pn collisions ($\sqrt s = 40$~GeV)
       from Ref. \protect \cite{DYGehrStir} confronted
       to the projected statistical errors expected for HERA-$\vec{N}$. 
       b) the acceptance for lepton pair registration in a particular detector
       with minimal registration angle $\Theta_{min}$.
       d) the unpolarized cross-section of the Drell-Yan process for $pp$ and
       $pn$ collisions \protect \cite{DYGehrStir}.
 }
\label{dygehr}
\vspace*{3mm}
\end{figure}

\begin{figure}[hbt]
\centering
\begin{minipage}[c]{7.5cm}
\centering
\epsfig{file=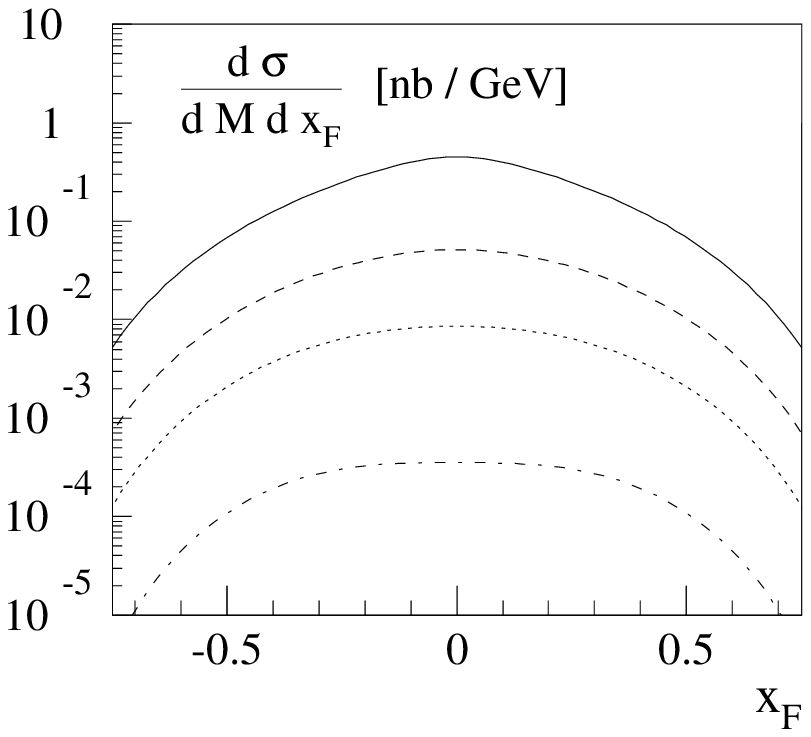,width=7.5cm}
\end{minipage}
\begin{minipage}[c]{7.5cm}
\centering
\epsfig{file=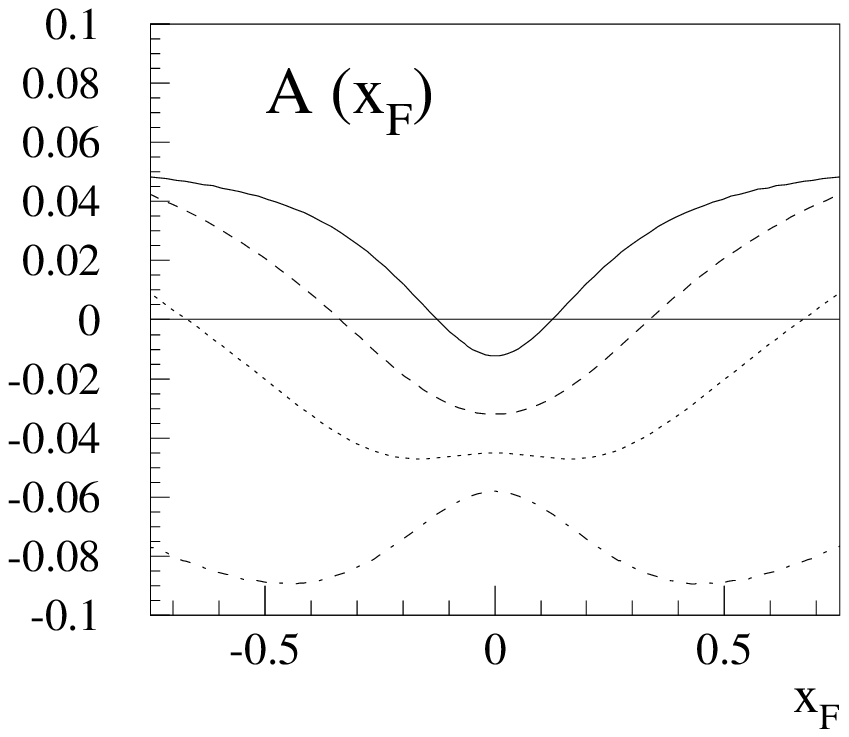,width=7.5cm}
\end{minipage}
\caption{\it The unpolarized Drell-Yan cross-section and the asymmetry in pp collisions
       for the polarized GS(A) parton distributions. Curves are for different
       invariant masses M: solid - 4 GeV; dashed - 6 GeV; dotted - 8 GeV;
       dot-dashed - 12 GeV. The figures are from Ref. \protect \cite{DYGehr}).
         }
   \label{xfdygehr}
\end{figure}

\vspace*{+0.5ex}
The prospects for such a measurement at HERA-$\vec N$
were calculated  in ref.~\cite{DYGehrStir} at next-to-leading order QCD. 
The spread of the predictions (see fig.~\ref{dygehr}a,c) reflects 
the insufficient present know\-ledge on the polarized sea quark
distributions in the region $x > 0.1$; not even the sign of the
asymmetry at large $M$ is predicted. Since the asymmetry is the weighted 
sum of $\Delta \bar u$ and $\Delta \bar d$ quarks with the strange
quark contribution assumed to be small and the weight of $\Delta \bar u$ 
is higher than that of $\Delta \bar d$ due to its abundance in the
proton and the electric charge, the asymmetry measured in $pp$ collisions
provides mainly 
information on $\Delta \bar u$, i.e. on the $u$ sea quark polarization.
The flavour contributions are slightly different for $pn$ collisions;
this results in an asymmetry being much smaller (fig.~\ref{dygehr}c) than in the $pp$
case. Since the total unpolarized cross-sections for the Drell-Yan process
 in $pp$ and $pn$
collisons are practically the same (see fig.~\ref{dygehr}d)
much larger luminosity is required in $pn$ collisons to obtain a reasonable statistical
sensitivity. Nevertheless,
it is very important as it could be used to decompose the flavour
structure of the polarized sea which is practically unknown at present.\\
Also, with a larger luminosity more information could be obtained from measuring
the differential lepton pair distributions in dependence on $x_F$ or 
$\eta$ \cite{DYGehrStir,DYGehr}, the predictions for the $x_F$ dependence are
shown in fig.~\ref{xfdygehr}. \\
We note that the acceptance for lepton pair detection
was not taken into account in the calculations \cite{DYGehrStir} as it depends
on the particular detector. The acceptance (integrated over kinematical
parameters of produced pairs) depends mainly on the minimal
accepted lepton angle in the detector (see fig.~\ref{dygehr}b);
a value of about 50\% may be realistic. In this case the projected
statistical sensitivity values, shown in fig.~\ref{dygehr}a, would be larger
by a factor of $\sqrt 2$. \\

\vspace*{+0.5ex}
{\bf Drell-Yan pair} production with {\bf transverse polarization} of
both beam and target can provide a measurement of the transversity
distribution, $\delta q(x)$.
The transverse double spin asymmetry in nucleon-nucleon
 Drell-Yan production 
can be schematically written in the form \cite{JaffeJi2}

\begin{figure}[hbt]
\centering
\epsfig{file=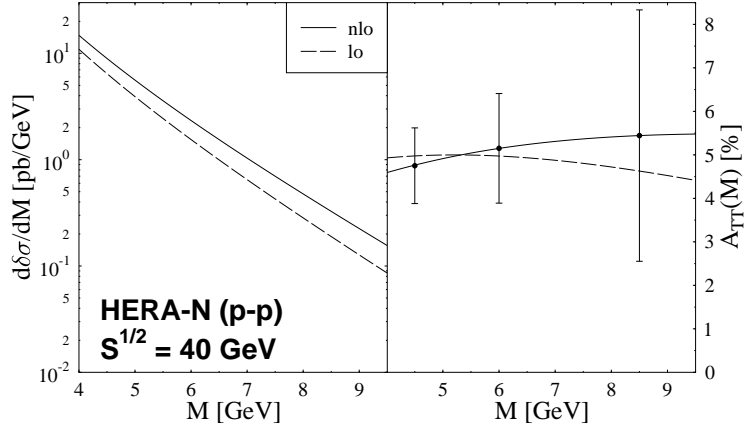,width=10cm}
\caption{\it Maximal polarized Drell-Yan cross-section (left figure) 
           and the asymmetry (right figure) in double transverse polarized collisions
           as a function of the invariant mass. The results are shown
           for both LO and NLO calculations \protect \cite{Martin}.
           The projected statistical errors for HERA-$\vec N$ are shown, as well.
            }
      \label{figdytt}
\end{figure}

\newpage

\begin{equation}
\label{DYTT}
A^{DY}_{TT} = {{\sin^2 \theta \cos{2 \phi}} \over {1 + \cos^2 \theta}}
{{\sum_i e^2_i [\delta q_i(x_1) \delta \bar q_i(x_2) + (1
    \leftrightarrow 2)]} \over 
{\sum_i e^2_i [q_i(x_1) \bar q_i(x_2) + (1
  \leftrightarrow 2)]} },
\end{equation}
\vspace*{+5mm}

where $\theta$ is the polar angle of one lepton in the virtual photon  
rest frame and $\phi$ is the angle between the direction of
polarization and the normal to the dilepton decay plane.
An estimate of the asymmetry for HERA-$\vec N$ energy was given recently
\cite{Martin} from both LO and NLO calculations.
One should stress, however, that the anticipated asymmetry level strongly 
depends on the actual size of the transversity distributions, which
are totally unknown at present. Although in the non-relativistic quark model 
the relation $\delta q(x) = \Delta q(x)$ holds, in reality differences
between both distributions are expected to be caused by dynamical effects.
Due to the lack of any information on the transversity distribution, the
maximally possible value of the asymmetry was estimated \cite{Martin};
the corresponding results on LO and NLO polarized Drell-Yan cross-section and asymmetry
are presented at fig.~\ref{figdytt}. The projected statistical errors for
a measurement of $A_{TT}$ at HERA-$\vec N$ are also shown. The maximal value
of $A_{TT}$  at an invariant mass of $M=4$ GeV was found to be approximately $4\%$
 with an expected statistical error 
of about $1\%$. The expected value of the asymmetry 
at RHIC energies is smaller but the statistical errors become relatively
smaller at $\sqrt{s}=500$ GeV due to the higher luminosity of ${\cal{L}}=800$~pb$^{-1}$
 \cite{Martin}. \\
The calculations \cite{Martin} do not account for the acceptance of the lepton pair
in the detector. The same discussion as above for the LL case applies here. \\

We note that there exist another potentially interesting 
possibility, the study of the {\bf longitudinal-transverse}
double spin asymmetry, $A^{DY}_{LT}$. 
This asymmetry was calculated in
ref.~\cite{JaffeJi2} and depends in a rather complicated fashion
on both twist-2 ($\Delta
q(x)$ and $\delta q(x)$) and twist-3 ($g_T(x)$ and $h_L(x)$) polarized
structure functions. In contrast to $A^{DY}_{LL}$ and $A^{DY}_{TT}$
the asymmetry $A^{DY}_{LT}$ decreases as $M/\sqrt{Q^2}$. 
The expected level of the asymmetry $A^{DY}_{LT}$ at HERA-$\vec N$ energy 
has not been calculated yet, the expected level of sensitivity
as a function of the lepton pair mass can be taken from fig.~\ref{dygehr}. \\




\section{Conclusions}

\vspace{1mm}
\noindent
A physics programme for a possible fixed target polarized nucleon-nucleon 
collision experiment utilizing 
an internal target in the 820 GeV HERA proton beam has been presented.
Single spin asymmetries, accessible already with the existing unpolarized 
beam, are found to be a powerful tool to study the nature
and physical origin of higher twist effects and a possible manifestation of
non-perturbative dynamics.
For that the study of asymmetries over a sufficiently large $p_T$-range
is essential; HERA-$\vec N$ would be able to provide data up to $p_T=$10~GeV/c
in the central region and up to 5$\div$6~GeV/c in the fragmentation region
of the polarized nucleon.
When measuring the polarized gluon distribution through double spin 
asymmetries in {\it photon (plus jet)} and {\it $J/\psi$ (plus jet)} 
production -- requiring a polarized HERA proton beam -- the projected 
statistical accuracies are found to be comparable to those predicted for the 
spin physics program at RHIC. 
Although both measurements explore the same $x_{gluon}$ range they are
complementary due to the different $p_T$ ranges accessible.
A measurement of
Drell-Yan pair production with both beam and target longitudinally
polarized can improve our knowledge on the polarized light sea quark 
distributions. A study of double transverse 
and/or longitudinal-transverse Drell-Yan spin asymmetries 
as well as a study of the single spin asymmetry in inclusive two-pion production
might open first access to the quark transversity distribution.
The existence of a polarized internal gas target in HERA-$\vec{N}$ would
allow to study polarized $pn$ and $pA( D, ^3He, ...)$
collisions, which are harder to be realized at RHIC.
In addition, there is a potential to obtain significant results on
the long-standing unexplained spin asymmetries in elastic scattering.

\section*{Acknowledgements}
%

\vspace{1mm}
\noindent
We are indebted to M.~Anselmino, O.~Teryaev and A.~Tkabladze
for very valuable comments. We thank T.~Gehrmann, O.~Martin, and W.~Vogelsang
for having us supplied with Drell-Yan predictions for HERA-$\vec N$ prior to
publication.


\end{document}